\documentclass[journal=jpcbfk,manuscript=article]{achemso}
\usepackage[utf8]{inputenc}

\usepackage{longtable}
\usepackage{multirow}
\usepackage{amsmath}
\usepackage[usenames,dvipsnames]{color}
\usepackage{soul}
\usepackage{threeparttable}
\usepackage{graphicx}
\usepackage{amsmath,amssymb}
\usepackage{caption}
\usepackage{color}
\usepackage{xcolor}
\usepackage{dcolumn}
\usepackage{bm}
\usepackage{float}
\usepackage{subcaption}
\usepackage{textcomp}
\usepackage{url}
\usepackage[normalem]{ulem}
\usepackage{booktabs}

\usepackage{microtype}

\usepackage{xr}
\makeatletter

\newcommand*{\addFileDependency}[1]{
\typeout{(#1)}

\@addtofilelist{#1}
\IfFileExists{#1}{}{\typeout{No file #1.}}
}\makeatother

\newcommand*{\myexternaldocument}[1]{%
\externaldocument{#1}%
\addFileDependency{#1.tex}%
\addFileDependency{#1.aux}%
}

\myexternaldocument{si}

\author{Cangtao Yin, Meenu Upadhyay} \affiliation[University of
  Basel]{Department of Chemistry, University of Basel,
  Klingelbergstrasse 80 , CH-4056 Basel, Switzerland}

\author{Markus Meuwly} \affiliation[University of Basel]{Department of
  Chemistry, University of Basel, Klingelbergstrasse 80 , CH-4056
  Basel, Switzerland} \email{m.meuwly@unibas.ch}

\title{Structure and Spectroscopy of Criegee Intermediates in Gas- and
  Aqueous Environments}

\begin{document}
\date{\today}

\begin{abstract}
The dynamics and spectroscopy of the small (H$_2$COO) and large
(CH$_3$CHOO) Criegee intermediates (CIs) in the gas phase, inside/on
water droplets, on amorphous solid water (ASW) and in bulk water are
investigated using validated energy functions. For both species,
facile diffusion between surface and inside positions for water
droplets are found whereas on amorphous solid water at low
temperatures (50 K) no surface diffusion is observed on the
multiple-nanosecond time scale. This is at variance with other
species, such as CO or NO on ASW. The infrared spectroscopy of both
CIs in contact with an aqueous environment leads to shifts of the
spectral features on the order of a few to a few tens of cm$^{-1}$,
depending on the vibrational mode considered. This is consistent with
Stark-induced spectral shifts for small molecules in protein
environments. However, the spectroscopy of both CIs in contact with
water droplets does not depend on the positioning relative to the
droplet (inside vs. surface).
\end{abstract}

\section{Introduction}
The nature and phase of chemical environments can play decisive roles
for the structure, spectroscopy, reactivity and other relevant
chemical properties of solutes. A particularly notable example
constitute chemical reactions on the surface of droplets which has
been a hotly debated subject over the past decade. Another area where
specific effects can be expected concerns the spectroscopic response
of solutes in different chemical environments. This arises in
atmospheric and interstellar chemistry where adsorbates experience
varying environments depending on the local properties of the
medium. For example, adsorbates on amorphous solid water have been
found to follow non-Arrhenius diffusion due to surface
roughness.\cite{MM.oxy:2018}\\

\noindent
Infrared (IR) and vibrational spectroscopy is a powerful
non-destructive technique to characterize systems in the gas- and
condensed phase. For example, IR spectra can provide information to
infer structure from spectroscopic responses. This has, for example,
been used to determine the metastable structure of free carbon
monoxide in the active site of
myoglobin.\cite{anfinrud:1995,anfinrud:1995.2,anfinrud:1997,anfinrud:1999,MM.mbco:2003,MM.mbco:2004,MM.mbco:2008,MM.mbco:2010}
Due to the strong inhomogeneous electric fields inside a protein,
spectral responses shift and split as a consequence of changes in the
structure. This has also been used to relate ligand binding strength
and frequencies of a spectroscopic reporter such as cyanolated
benzenes.\cite{boxer:2006,MM.stark:2017}\\

\noindent
Aqueous (micro)droplets and aerosols are environments relevant to
atmospheric chemistry and laboratory-based experiments.  The strong
electric fields at the interfaces of aqueous droplets can influence
the spectroscopy of guest molecules and their
reactivity.\cite{zare:2020,yan:2016,zare:2017,wei:2020}
Spectroscopically, stimulated Raman excited fluorescence microscopy
has been used to measure a 5 cm$^{-1}$ red shift for the -CN stretch
vibration in rhodamine 800 between the bulk and the microdroplet
environment.\cite{zare:2020} In addition, chemical reactions in and on
water droplets have recently been found to be accelerated by orders of
magnitudes compared with the same reactions in bulk
water.\cite{yan:2016,zare:2017,wei:2020} How exactly microdroplets
lead to such accelerations at a molecular level is still debated but
it is suspected that a number of factors contribute to this
observation: solvent evaporation from the droplet, confinement of the
reagents, the electric fields ($10^7$ V/cm, surface potential 3.63
V)\cite{kappes:2021} at the air-solvent interface, droplet size, or
molecular orientation at the interface to name a few. On the other
hand, for a Diels-Alder reaction that occurs readily in bulk solvent
it was found that no product was formed at all.\cite{zare:2017} This
is consistent with work on pyruvic acid which reported different
chemistry depending on whether the photochemistry occurred in aqueous
bulk environment or at the air-water interface.\cite{kappes:2021}\\

\noindent
Criegee intermediates (CIs), generated during alkene ozonolysis, are
highly reactive carbonyl oxides that play crucial roles in atmospheric
oxidation, notably in the conversion of SO$_2$ to H$_2$SO$_4$ and the
formation of secondary organic aerosols (SOA).\cite{ChhantyalPun2017}
Generally speaking, CIs are generated from alkene ozonolysis through a
1,3-cycloaddition of ozone across the C=C bond to form a primary
ozonide which then decomposes into carbonyl compounds and energized
carbonyl oxides, known as CIs.\cite{criegee1949ozonisierung} Such
highly energized intermediates rapidly undergo either unimolecular
decay to hydroxyl radicals\cite{alam2011total} or collisional
stabilization\cite{novelli2014direct}. Stabilized Criegee
intermediates can isomerize and decompose into products including the
OH radical, or undergo bimolecular reactions with water vapor, SO$_2$,
NO$_2$ and acids.\cite{taatjes2017criegee,mauldin2012new}\\

\noindent
Aerosols can substantially influence CI chemistry by providing
surfaces and aqueous microenvironments that promote adsorption,
stabilization, or enhanced reactions with water, SO$_2$ and
organics.\cite{Mao2013,Kuwata2020} Such heterogeneous or interfacial
processes can alter CI lifetimes and branching ratios, affecting
radical budgets and oxidation capacity. For instance, solvation or
hydrogen bonding within droplets may accelerate CI hydration or
rearrangement to peroxides, while reactions at aerosol surfaces may
enhance sulfuric acid formation and SOA growth.\cite{Sheps2017}
Consequently, aerosols act not only as sinks but as active
participants that modulate Criegee intermediate reactivity and their
broader atmospheric impacts.\\

\noindent
The present work focuses on the adsorption and spectroscopic
properties of the two smallest CIs: H$_2$COO and {\it
  syn-}CH$_3$CHOO. First, the methods used are described. This is
followed by results on the structural characterization of the CIs in
the various environments. Next, the infrared spectra of the two CIs
depending on the environment are analyzed. Finally, conclusions are
drawn.\\

\section{Methods}
\subsection{Molecular Dynamics Simulations}
All molecular dynamics (MD) simulations were carried out with the
CHARMM and pyCHARMM suite of
codes.\cite{Charmm-Brooks-2009,pycharmm:2023,MM.pycharmm:2023,MM.charmm:2024}
For both systems studied here, H$_2$COO and {\it syn}-CH$_3$CHOO, two
different and validated reactive PESs from previous work are
available.\cite{MM.criegee:2021,MM.criegee:2023,MM.h2coo:2024,MM.h2coo:2025}
The PESs are based on multi-state adiabatic reactive MD
(MS-ARMD)\cite{MM.msarmd:2014} and a neural network-based
representation for which the PhysNet architecture was
used.\cite{MM.physnet:2019}\\

\noindent
Simulations using MS-ARMD were run using CHARMM, whereas for the mixed
machine learning/molecular mechanics (ML/MM) MD simulations the
pyCHARMM code was employed. In all simulations the systems were first
heated and equilibrated at the target temperature before production
simulations, 2 ns in length, were carried out. Four different system
setups were considered: the CIs in the gas phase, CIs adsorbed on
amorphous solid water (ASW), CIs on and within water droplets, and
finally CIs in bulk water. In all simulations the time step was
$\Delta t = 0.1$ fs.\\

\noindent
For CIs adsorbed on water a previously generated setup for the ASW was
used.\cite{MM.oxy:2014,MM.oxy:2018} The initial ASW structure was
generated as described previously.\cite{MM.oxy:2014} Starting from a
TIP3P\cite{Jorgensen.tip3p:1983} water box ($31 \times 31 \times 50$
\AA\/$^3$), equilibrated at 300 K, the system is first quenched to 50
K, then equilibrated in the \textit{NpT} ensemble, followed by further
equilibration with \textit{NVT} using periodic boundary
conditions. Next, the CIs were adsorbed onto different locations of
the surface, their structure was relaxed and heated to and
equilibrated at 50 K after which production simulations followed.\\

\noindent
All simulation environments used the flexible KKY model for
water.\cite{KKY:1994} The implementation of the KKY model in CHARMM
was that used in previous work.\cite{MM.ice:2008,MM.o2:2019} Droplets
reminiscent of aerosol particles were generated by starting from a
cubic box of water containing 452 water molecules. Stochastic boundary
conditions were applied to generate a droplet and for the duration of
the simulations, water molecules were hindered to escape from the
droplet by applying a center-of-mass constraint. These simulations
used the KKY water model throughout. After preparation, the CIs were
placed on top of the droplet surface. Heating and equilibration were
carried out with a target temperature of 300 K. Following this, $NVE$
simulations were carried out and the CIs remained unconstrained. In
other words, the adsorbates were free to remain on top of the droplet,
to diffuse into the droplet or to disengage from it.\\

\noindent
For the simulations in solution, a solvent box with dimensions $40
\times 40 \times 40$ \AA\/$^3$ was generated using Packmol. The solute
was placed into the center of the simulation box and remained there
through application of a mild harmonic restraint ($k = 1$
kcal/mol/\AA\/$^2$) applied to the center-of-mass. Heating,
equilibration and production dynamics were carried out at 300 K in the
$NVT$ ensemble. The long-range interactions were. The non-bonded
forces between atoms separated by within 14 \AA\/ were computed, but
smoothly switches them off between 10 and 12 \AA\/.\\

\noindent
MS-ARMD is a computationally efficient means to investigate chemical
reactions. The force fields for the reactant and products were
separately parameterized and the Gaussian and polynomial (GAPO)
functions \cite{MM.msarmd:2014} were used to connect the reactant and
product force fields to generate a continuous reactive PES along the
reaction paths. On the other hand, PhysNet is a neural network that
uses energies, forces, charges and dipole moments to minimize a loss
function for obtaining a trained ML-PES. For
H$_2$COO\cite{MM.h2coo:2024,MM.h2coo:2025} the CCSD(T)-F12a/aVTZ and
CASPT2/aVTZ level of theories were used for calculating the reference
datasets for fitting the MS-ARMD and PhysNet models, respectively,
whereas for {\it syn}-CH$_3$CHOO, MP2/6-311++G(2d,2p) and
CASPT2(12,10)/cc-pVDZ were used.\cite{MM.criegee:2023}\\

\subsection{Analysis}
Structural and spectroscopic properties of the systems and the CIs
were analyzed. The radial distribution functions $g(r)$ were
determined by using VMD.\cite{vmd} The IR spectra $I(\omega)$ were
obtained via the Fourier transform of the dipole-dipole correlation
function from the dipole moment time series
\begin{equation}
  I(\omega) n(\omega) \propto Q(\omega) \cdot \mathrm{Im}\int_0^\infty
  dt\, e^{i\omega t} 
  \sum_{i=x,y,z} \left \langle \boldsymbol{\mu}_{i}(t)
  \cdot {\boldsymbol{\mu}_{i}}(0) \right \rangle
\label{eq:IR}
\end{equation}
A quantum correction factor $Q(\omega) = \omega(1-exp(-\beta \hbar
\omega))$ was applied to the results of the Fourier
transform.\cite{marx:2004}\\

\section{Results}
\begin{figure} [H]
    \centering
    \includegraphics[width=0.5\linewidth]{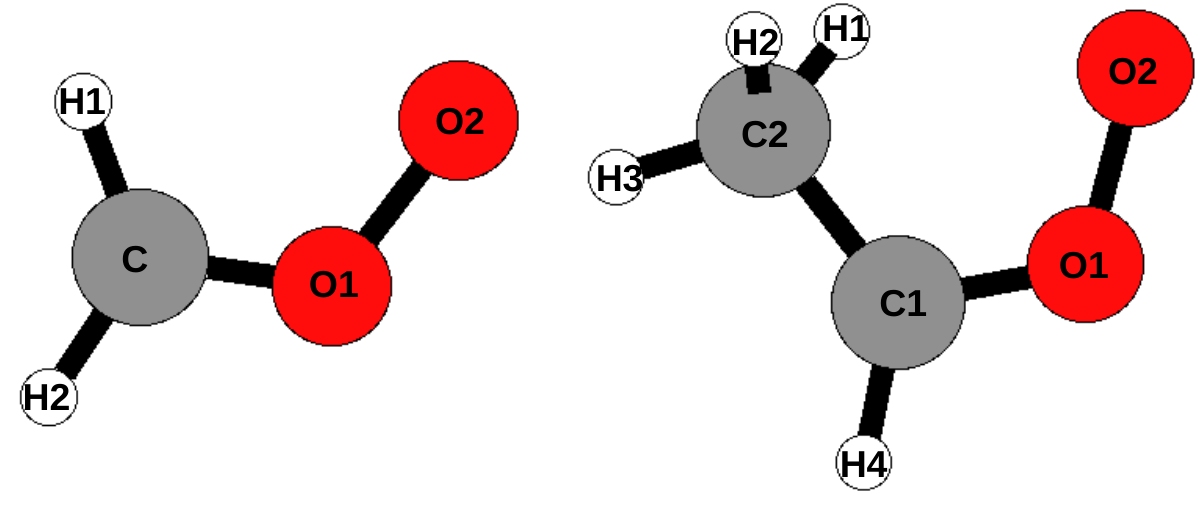}
    \caption{The structures of two Criegee intermediates, H$_2$COO and
      {\it syn}-CH$_3$CHOO, with atoms labelled.}
    \label{fig:structures}
\end{figure}

The structures of the two Criegee intermediates considered here are
shown in Figure \ref{fig:structures}. For both CIs, H$_2$COO and {\it
  syn}-CH$_3$CHOO, the IR spectra were studied experimentally in the
gas phase.\cite{lee:2013,lee:2015} However, no data is available for
the spectroscopy in solution, although some computational studies
focused on the stability \cite{francisco:2017} and new reaction
mechanisms which leads to a new product HOCH$_2$OOH in bulk
water.\cite{francisco:2016}.\\

\noindent
The PESs used in the present work were previously validated vis-a-vis
experimental measurements. For the ``small CI" (s-CI, H$_2$COO) this
included the vibrational spectrum,\cite{lee:2013} the lifetime of the
activated molecule and the fragmentation
channels.\cite{stanton:2015,lester:2024} In particular, the recently
trained ML-PES\cite{MM.h2coo:2025} successfully identified the HCO+OH
dissociation product as the ``minority channel". Also, the computed
vibrational frequencies including anharmonic corrections and
mode-coupling are in reasonably good agreement with the experimental
data. One difficulty in directly comparing measured and computed
frequencies is in the challenging experimental characterization. The
IR spectra for both, s-CI and l-CI, were not determined on pure
samples but needed to be extracted from spectroscopic measurements of
a reaction mixture through subtraction of other species' spectra,
including CH$_2$I$_2$, N$_2$, O$_2$, dioxirane, methylenebis(oxy), and
{\it cis}-CH$_2$IOO.\cite{lee:2013,lee:2015} Table
\ref{sitab:frequencies_h2coo} compares computed harmonic frequencies
with experimentally assigned data for H$_2$COO.\cite{lee:2013} For the
framework modes (below 2000 cm$^{-1}$), anharmonic corrections are
estimated to be on the $\sim 10$ to $\sim 50$ cm$^{-1}$ depending on
the mode considered.\cite{MM.h2coo:2025,lester:2024}\\

\noindent
For the larger (l-CI, {\it syn-}Criegee) the energy-dependent rates
and final vibrational and rotational state distributions of the
OH-fragment were in excellent agreement with
measurements.\cite{lester:2016,MM.criegee:2021,MM.criegee:2023} For
the computed vibrational frequencies compared with measurements the
same remarks as for the s-CI above apply. Table
\ref{sitab:frequencies_ch3choo} indicates that the computed harmonic
frequencies using the ML-PES are in acceptable agreement with those
reported from the experiments. Including anharmonic corrections will
improve this comparison but is not done in the present work also
because the data to compare with from experiment has some
uncertainty. Hence, for both systems vetted MS-ARMD- and ML-PESs are
available and used in the following.\\

\subsection{Structure of the Water Environment}
\begin{figure} [H]
    \centering
    \includegraphics[width=1.0\linewidth]{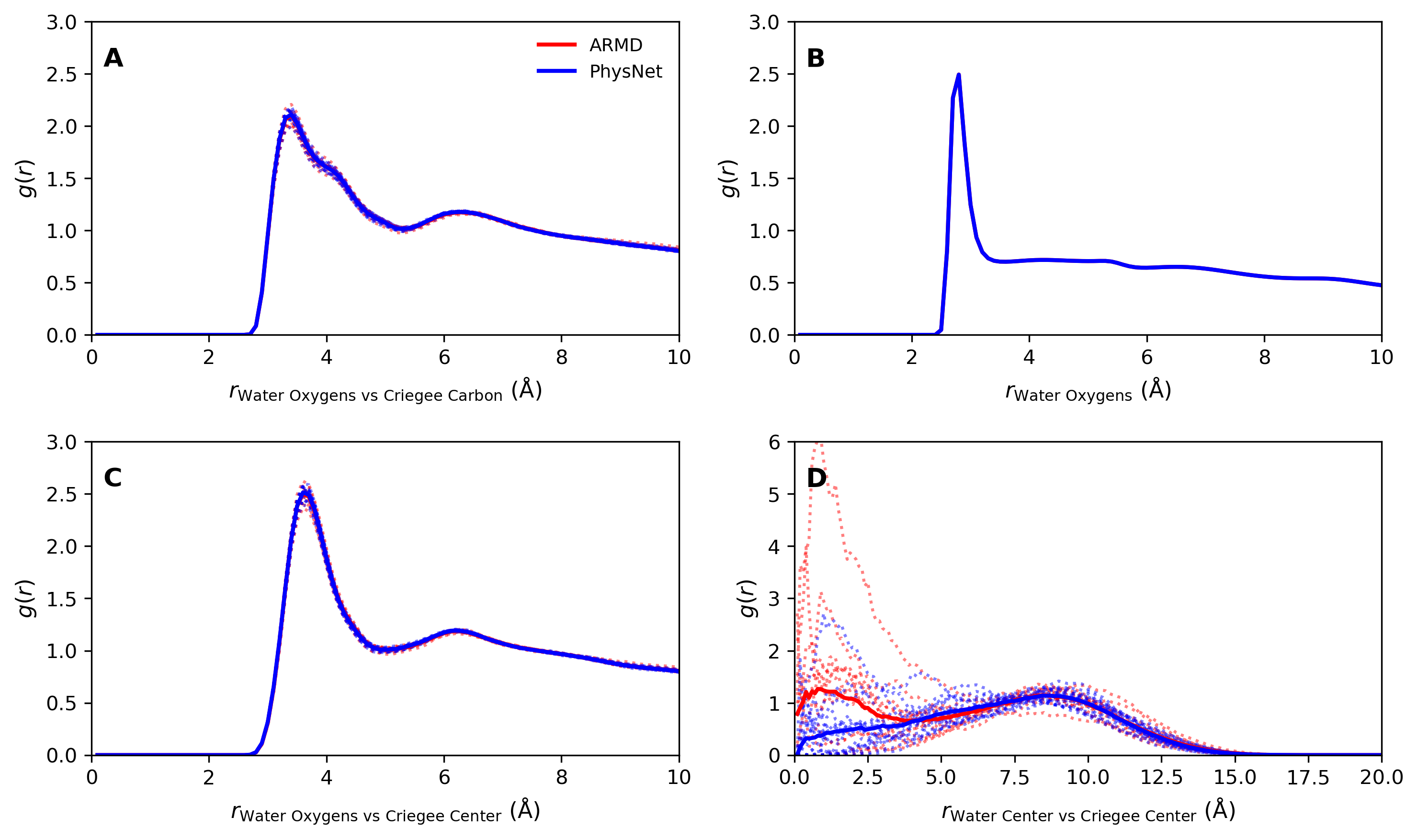}
    \caption{Radial pair distribution function for H$_2$COO (CI) on
      the water droplets. The radial distances in panels A to D are
      the O$_{\rm W}$--C$_{\rm CI}$, O$_{\rm W}$--O$_{\rm W}$, O$_{\rm
        W}$--CoM$_{\rm CI}$, and CoM$_{\rm droplet}$--CoM$_{\rm CI}$
      separations. Averages over 20 independent simulations, each 2 ns
      in length, are shown for MS-ARMD (red thick lines) and PhysNet
      (blue thick lines) energy functions for H$_2$COO,
      respectively. Radial distribution functions for individual
      trajectories are shown as dotted lines. The two extreme cases,
      CI located inside the water droplet and adsorbed on its surface
      are shown in Figure \ref{sifig:droplet}.}
    \label{fig:c1-gofr-drop}
\end{figure}

Next, the structure of the solvent molecules around the two CIs
together with the water droplets was characterized. Figure
\ref{fig:c1-gofr-drop} reports various radial distribution functions
for H$_2$COO adsorbed to and within water droplets. Solid and dashed
lines refer to the average and 20 individual MD trajectories run with
the MS-ARMD (red) and ML-PES (blue) energy functions,
respectively. The water structuring around C$_{\rm CI}$ (panel A), the
droplet structure (panel B), and the water structure around the center
of mass of H$_2$COO (panel C) are virtually identical for both
PESs. On the other hand, using the center of mass separation between
the droplet and the guest molecule, some differences are found (panel
D). Towards the droplet surface ($r_{\rm CoM_{Water}-CoM_{CI}} > 7.5$
\AA\/) the structures are comparable. The radial pair distributions
for O$_{\rm W}$--C$_{\rm CI}$, O$_{\rm W}$--O$_{\rm W}$, O$_{\rm
  W}$--CoM$_{\rm CI}$ are well converged for each individual
simulation, giving nearly identical results (Panels A–C). However,
when using the relative positioning of the centers of mass of the
droplet and the CI (CoM$_{\rm droplet}$--CoM$_{\rm CI}$), the scatter
is wider and for a converged average a larger number of independent
simulations will be required. The individual $g(r)$ in Figure
\ref{fig:c1-gofr-drop}D indicate that the CI can move between inside
and surface sites. Figure \ref{sifig:droplet} reports individual
radial distribution functions for an inside position (using the
MS-ARMD PES) and a surface position (using the ML-PES). It is found
that the droplet structure is nearly identical (panel B), whereas the
other three distribution functions differ to smaller or greater
extents (panels A, C, D).\\

\begin{figure} [H]
    \centering \includegraphics[width=1.0\linewidth]{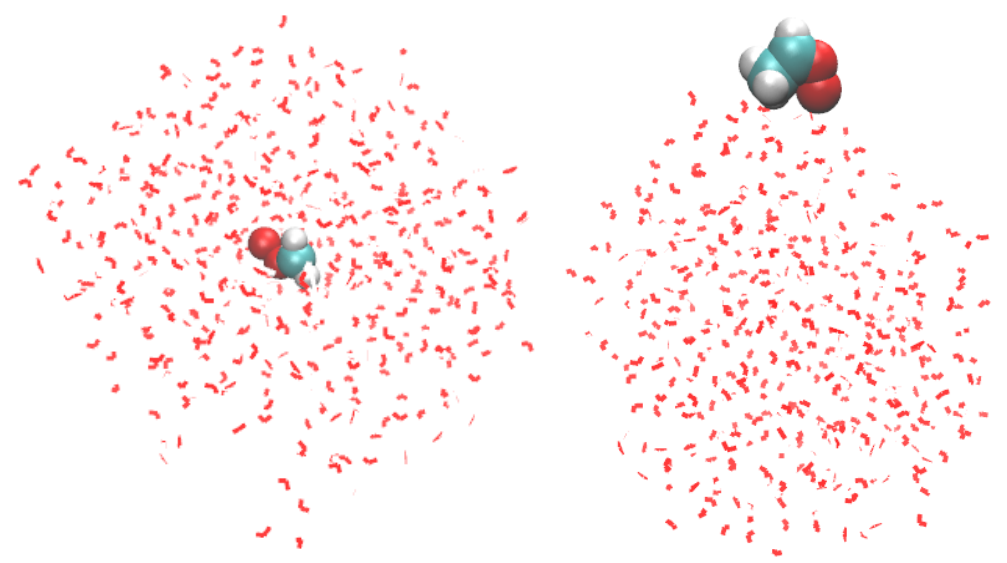}
    \caption{Snapshots of the droplet simulations with H$_2$COO (left)
      and {\it syn}-CH$_3$CHOO (right) as the guest molecule.}
    \label{fig:snapshot-drop}
\end{figure}

\noindent
Figure \ref{fig:snapshot-drop} reports two structures for H$_2$COO and
{\it syn}-CH$_3$CHOO attached to the droplet surface. It is found that
the guest molecule can fully insert despite the apolar
CH$_2$-group. The partial atomic charges used in the MS-ARMD energy
function are: C(0.313$e$), H1(0.205$e$), H2(0.192$e$), O1(--0.174$e$),
and O2(--0.536$e$) (see Figure \ref{fig:structures} for
labels). Hence, the possibility for insertion is primarily driven by
the hydration of the two oxygen atoms which form hydrogen bonds with
the droplet-water molecules. Figure \ref{fig:snapshot-drop} also shows
that the water droplet is not always perfectly spherical.\\

\begin{figure} [H]
    \centering
    \includegraphics[width=1.0\linewidth]{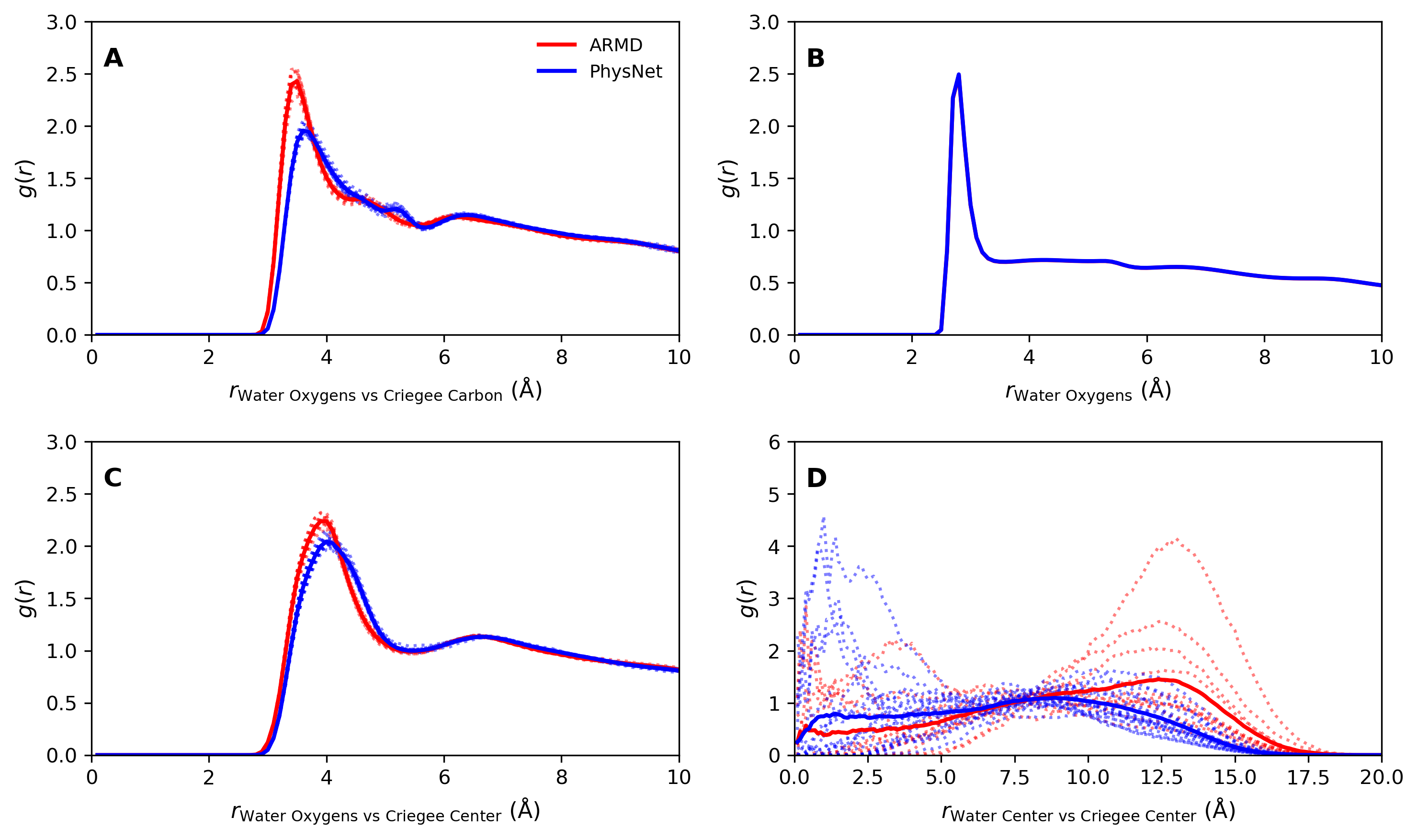}
    \caption{Radial pair distribution function for {\it
        syn}-CH$_3$CHOO (CI) on the water droplets. The radial
      distances in panels A to D are the O$_{\rm W}$--C$_{\rm CI}$,
      O$_{\rm W}$--O$_{\rm W}$, O$_{\rm W}$--CoM$_{\rm CI}$, and
      CoM$_{\rm droplet}$--CoM$_{\rm CI}$ separations. Averages over
      20 independent simulations, each 2 ns in length, are shown for
      MS-ARMD (red thick lines) and PhysNet (blue thick lines) energy
      functions for {\it syn}-CH$_3$CHOO, respectively. Radial
      distribution functions for individual trajectories are shown as
      dotted lines.}
    \label{fig:c2-gofr-drop}
\end{figure}

\noindent
For {\it syn}-CH$_3$CHOO (see Figure \ref{fig:c2-gofr-drop}), all
three radial distribution functions involving the CI differ (panels A,
C, D). The first peak for $g(r)$ with $r_{\rm O_W-C_{CI}}$ is more
intense if the MS-ARMD energy function is used in the simulations. For
the O$_{\rm W}$--CoM$_{\rm CI}$ separation as the relevant coordinate
(Figure \ref{fig:c2-gofr-drop}C) the radial distribution functions are
closely overlapping whereas for the $r_{\rm CoM_{Water}-CoM_{CI}}$
differences are largest. With the MS-ARMD energy function (red
traces), {\it syn}-CH$_3$CHOO prefers to reside near the droplet
surface whereas using the ML-PES, $g(r)$ plateaus up to $r \sim 10$
\AA\/ after which it decays (blue traces). Considering individual
trajectories (dotted lines), the differences between the two energy
functions become even more apparent.\\

\begin{figure} [H]
    \centering
    \includegraphics[width=1.0\linewidth]{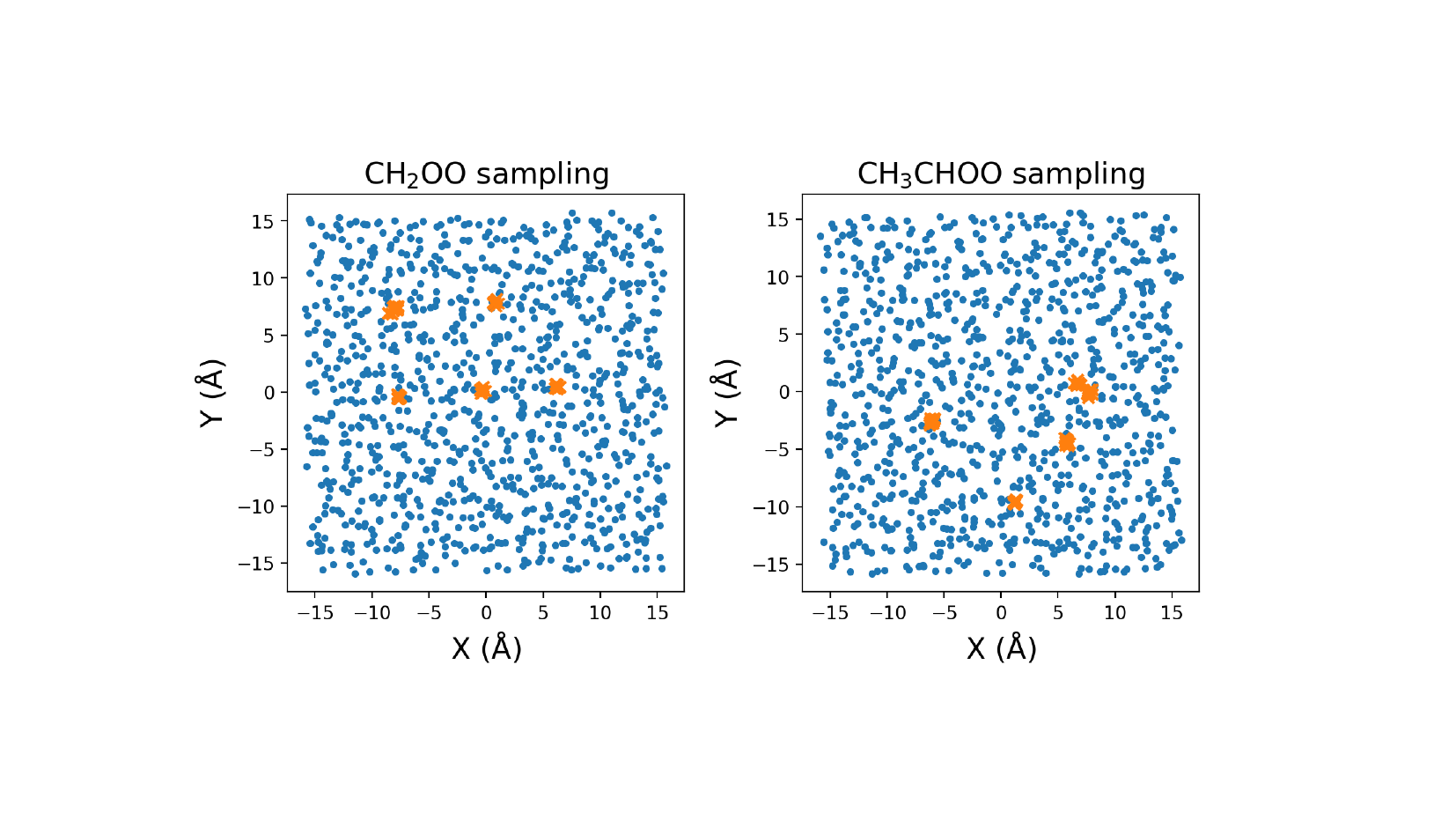}
    \caption{The CoM of CI sampled from 2 ns MD simulations at 50
      K. The water molecules are free to move but only oscillate
      around their positions at such low temperature. The CIs are free
      to move but do not diffuse between neighboring wells at the low
      temperatures on short time scales.}
    \label{fig:ci-position-asw}
\end{figure}

\noindent
For both CIs on the surface of ASW the motional freedom is rather
restricted due to the low temperature at which the simulations were
carried out (50 K). The CoM of both CIs remains close to its initial
position on the 2 ns time scale although diffusion to neighboring
sites would be possible in principle. Typical locations for the
CoM$_{\rm CI}$ are indicated as orange point clouds in Figure
\ref{fig:ci-position-asw}. Surface diffusion for atoms and diatomic
molecules involves barriers up to a few kcal/mol. As an example,
experimentally and computationally estimated diffusion barriers for CO
on ASW ranged from 0.25 to 1.85
kcal/mol.\cite{cuppen:2014,kouchi:2020,acharyya:2022,MM.asw:2024}
Evidently, these diffusional barriers are higher for the two CIs.\\

\subsection{Infrared Spectroscopy of CIs in Different Environments}
\begin{figure}[h!]
    \centering
    \includegraphics[width=0.7\linewidth]{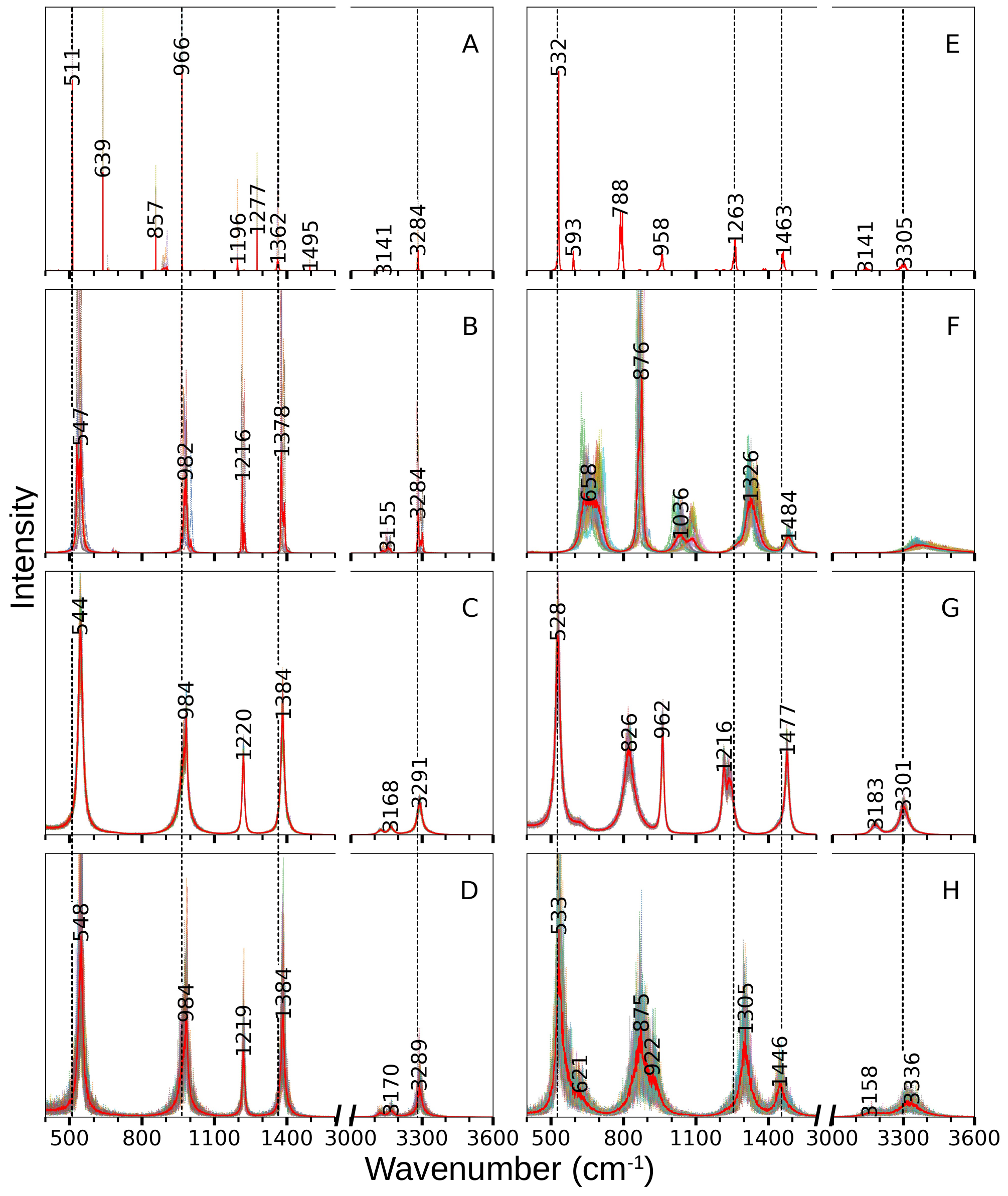}
    \caption{Simulated IR spectrum of H$_2$COO in four
      environments. Spectra from individual simulations and their
      average are the dotted and solid lines, respectively.  for each
      case were. Pronounced peaks in the averaged spectra are labelled
      with the respective wavenumber. Panels A to D for simulations
      using the MS-ARMD energy function. Panels E to H for simulations
      using the ML-PES. Panels A/E, B/F, C/G, and D/H for simulations
      in the gas phase, on ASW, on/in water droplets, and in bulk
      water.}
    \label{fig:c1-spec}
\end{figure}

\begin{figure}[h!]
    \centering
    \includegraphics[width=0.7\linewidth]{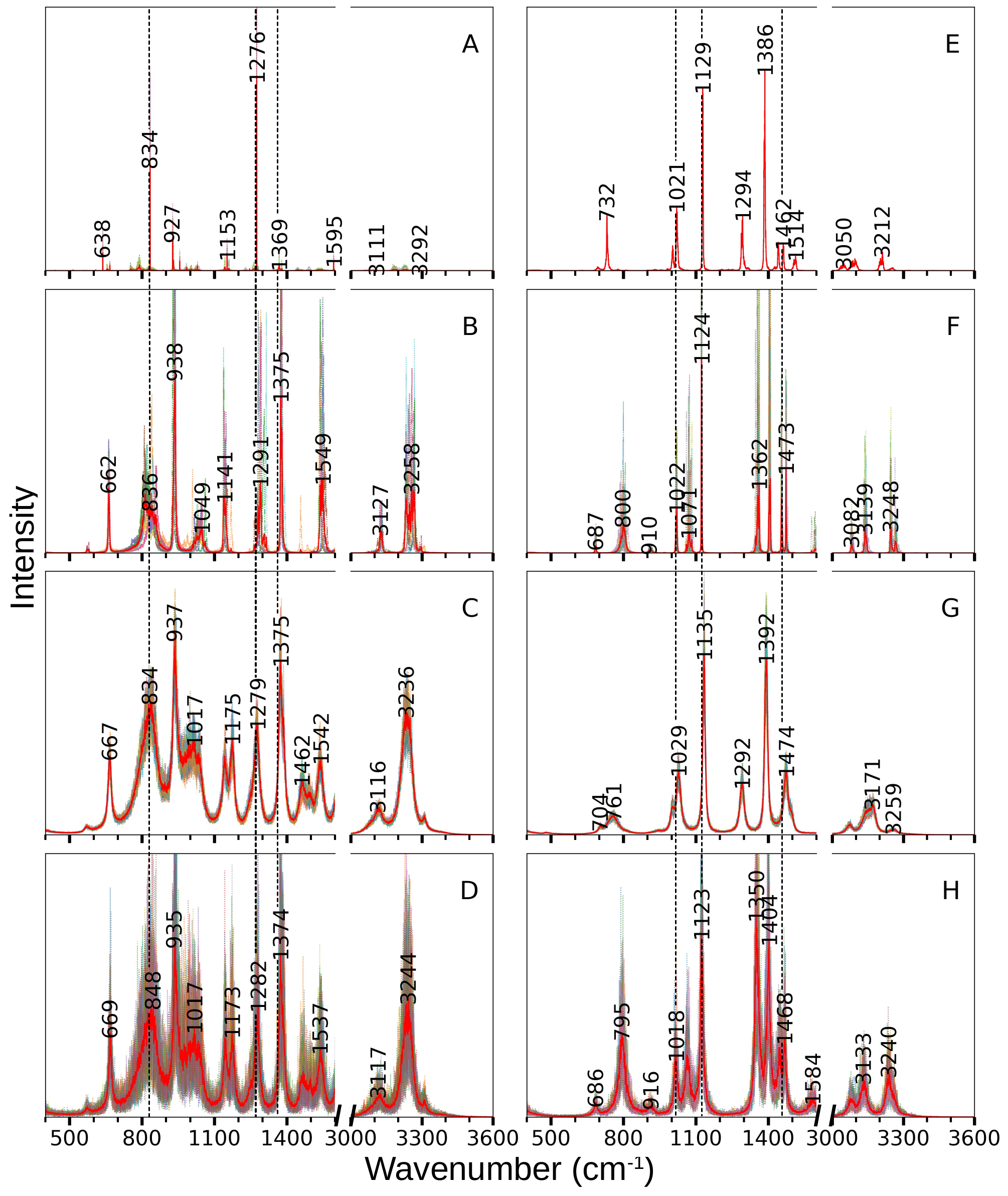}
    \caption{Simulated IR spectrum for {\it syn}-CH$_3$CHOO in four
      environments. The independent simulations for each case were
      presented with dotted lines and the average was presented with a
      red solid line. Pronounced peaks in the averaged spectra are
      labelled with the respective wavenumber. Panels A to D for
      simulations using the MS-ARMD energy function. Panels E to H for
      simulations using the ML-PES. Panels A/E, B/F, C/G, and D/H for
      simulations in the gas phase, on ASW, on/in water droplets, and
      in bulk water.}
    \label{fig:c2-spec}
\end{figure}

Next, the IR spectra for both CIs in the gas phase and the different
environments are presented and discussed, see Figures
\ref{fig:c1-spec} and \ref{fig:c2-spec}. As a reference, the averaged
gas phase IR spectrum was determined from 20 independent trajectories,
see panels A/E in Figures \ref{fig:c1-spec} and \ref{fig:c2-spec}. For
H$_2$COO three framework modes (below 2000 cm$^{-1}$) and one
high-frequency mode were selected, see dashed vertical lines in Figure
\ref{fig:c1-spec}. From left to right, each dashed vertical line
corresponds to the C-O-O scissor, O–O stretch, C-O stretch/H-C-H
scissor, asymmetric C–H stretch modes in the left panels and C-O-O
scissor, C-O stretch scissor, H-C-H scissor, asymmetric C–H stretch in
the right panels. Approximate mode assignments are also given in Table
\ref{sitab:frequencies_h2coo}.\\

\noindent
All modes experience more or less pronounced environment-induced
spectral shifts. Simulations using the MS-ARMD PES shift the mode at
511 cm$^{-1}$ by $\sim 30$ cm$^{-1}$ to the blue in panels B to
D. This is comparable to the 966 cm$^{-1}$ mode whereas the 1362
cm$^{-1}$ shifts rather by $\sim 20$ cm$^{-1}$ to the blue. The blue
shifts for the high-frequency CH-stretch mode is only $\sim 10$
cm$^{-1}$. For simulations using the ML-PES (panels E to H) frequency
shifts are observed as well. For this, the [532, 1263, 1463, 3305]
cm$^{-1}$ modes were considered. The patterns of frequency shifts
differs considerably from using the empirical MS-ARMD PES. Both, blue
and red shifts are observed and some motions do not occur at all. For
example, on ASW the 532 cm$^{-1}$ is entirely absent, whereas in/on
the water droplet and in bulk water the mode shifts to the red by $-4$
cm$^{-1}$ and to the blue by 1 cm$^{-1}$, respectively. Similar
observations are made for the other modes considered.\\

\noindent
For the l-CI the findings follow those for H$_2$COO in that all
environments lead to more or less pronounced frequency shifts relative
to the gas phase. Typically, using the MS-ARMD energy function leads
to blue shifts: the 834 cm$^{-1}$ signal from the gas phase appears at
[836, 834, 848] cm$^{-1}$ (ASW, droplet, bulk) whereas for the 1276
cm$^{-1}$ peak the different environments lead to [1291, 1279, 1282]
cm$^{-1}$ signatures which are all shifted to the blue, see Figure
\ref{fig:c2-spec}A/D. As for the s-CI, the ML-PES leads to blue- as
well as red-shifts. One example concerns the 1021 cm$^{-1}$ peak which
shifts to [1022, 1029, 1018] cm$^{-1}$.\\

\begin{figure}[h!]
    \centering
    \includegraphics[width=1.0\linewidth]{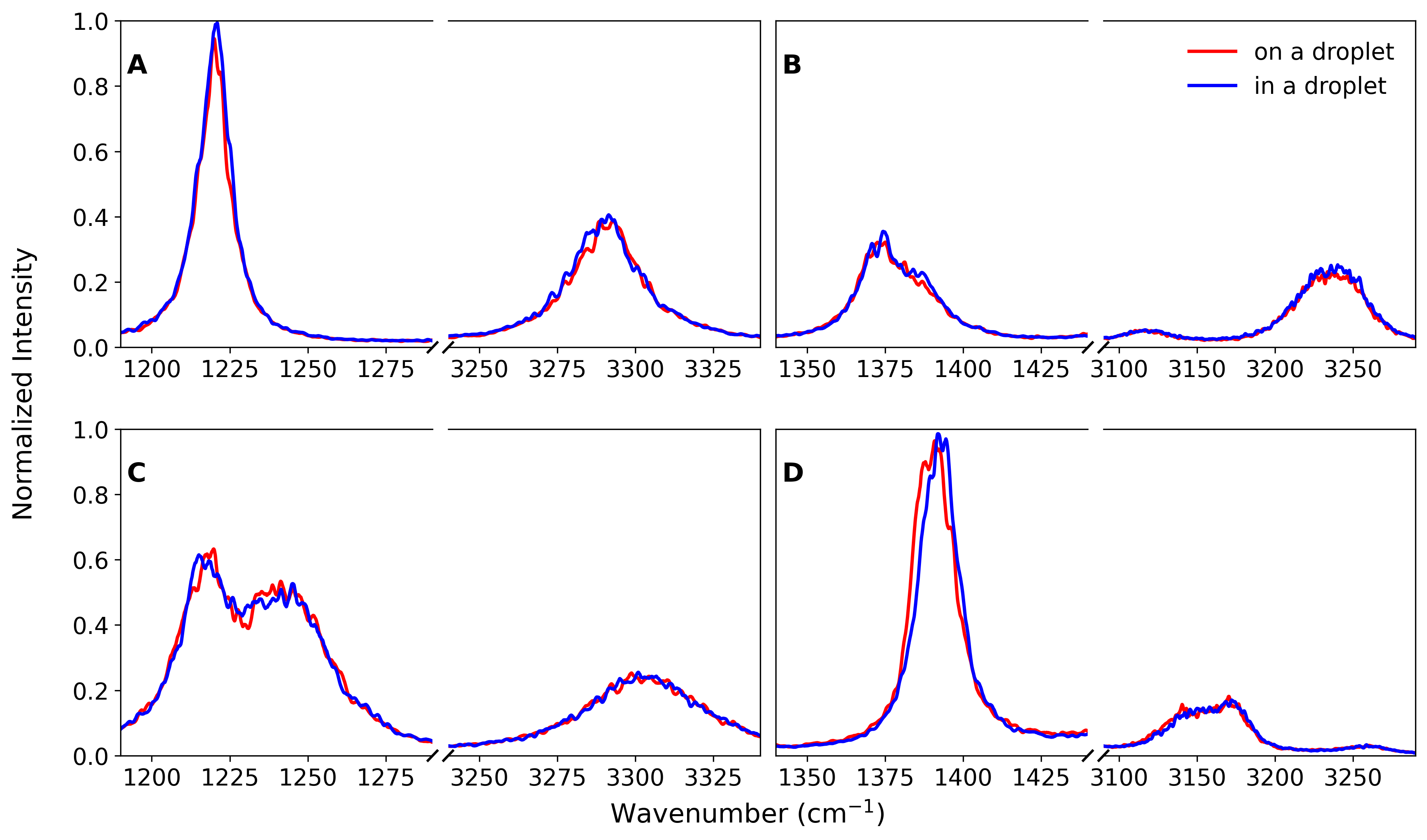}
    \caption{Comparison of IR spectrum with H$_2$COO (panels A and C)
      and {\it syn}-CH$_3$CHOO (panels B and D) on the droplet surface
      vs. inside the droplet. Top and bottom panels are from
      simulations using the MS-ARMD and ML-PES, respectively.}
    \label{fig:droplet-spec}
\end{figure}

\noindent
Since for the water droplet the s-CI and l-CI host molecules can
reside either on the surface or within the droplet, it is also of
interest to investigate whether the IR spectra between these two
situations differ. An overview of the spectra for both CIs is given in
Figures \ref{sifig:c1_on_in_full} and \ref{sifig:c2_on_in_full} using
both PESs. The difference spectra (green) indicate that for both guest
molecules the differences in the spectroscopy between adsorbed and
inserted species are small.\\

\noindent
For a more detailed analysis, individual spectral features were
selected, see Figure \ref{fig:droplet-spec}. Independent of the energy
function used - MS-ARMD (panels A, B) or PhysNet (panels C, D) - the
spectra with the s-CI or l-CI inside (blue) or on the surface (red) of
the droplet only differ marginally. For the l-CI, using the MS-ARMD
energy function, again no differences in the IR spectroscopy were
found (Figure \ref{fig:droplet-spec}B) whereas using the ML-PES the
framework mode around 1400 cm$^{-1}$ may respond slightly differently
to the l-CI residing on the droplet (red) compared with the guest
molecule being within the droplet (blue, shifted to higher frequency),
see Figure \ref{fig:droplet-spec}D. For the high-frequency CH-stretch
modes no differences between adsorbed and inserted guest molecule were
found. It is possible that these findings change for larger droplets
than those considered here, though.\\

\section{Discussion and Conclusions}
The present work considers the dynamics and spectroscopy of the two
smallest Criegee intermediates, H$_2$COO and {\it syn-}CH$_3$CHOO, in
the gas phase and in different hydration environments. For both
systems, validated energy functions based on MS-ARMD and ML-based
representations are available. The interest in environmental effects
on the dynamics and spectroscopy derives primarily from the relevance
of the two CIs in atmospheric chemistry where they can interact with
water through adsorption on water droplets or water-microcrystals.\\

\noindent
Most relevant for atmospheric chemistry are CIs in contact with water
droplets. The present simulations find that both CIs can reside on the
surface of droplets but also diffuse to the interior and stabilize
there. The IR spectroscopy is largely independent on whether the CIs
adsorb or insert. This is found for both PESs used in the present
work. For the dynamics on ASW at low temperatures (50 K), both CIs
remain at the initial adsorption sites.\\

\noindent
Measurements are available for the two molecules in the gas phase
whereas the spectroscopy of CIs together with water droplets or ASW
under laboratory conditions has not been reported so far. Typical
measured reaction rates for CIs with water range from $10^{-13}$
cm$^3$s$^{-1}$ to $10^{-16}$ cm$^3$molecule$^{-1}$s$^{-1}$ which
correspond to $\sim$seconds and $\sim$milliseconds depending on
relative humidity for water densities of $10^{17}$ cm$^3$ (high) and
$10^{16}$ cm$^3$ (low), respectively.\cite{taatjes:2013,liu:2025}
Hence, the lifetimes for CIs in contact with bulk water are
sufficiently long that IR spectra can be obtained from either
laboratory or earth-bound experiments. Computationally, the stability
of s-CI and l-CI with one and two water molecules was determined. They
range from $\sim 2$ kcal/mol to $\sim 10$ kcal/mol depending on the
quantum chemical methods
used.\cite{lin:2016,francisco:2016,francisco:2017} Computed effective
rates for the reaction of s-CI with two water molecules at 298 K and
20 \% humidity were $10^4$ s$^{-1}$.\cite{francisco:2016} Given these
findings it is anticipated that both, s-CI and l-CI, remain unreacted
on the time scale of the present simulations (multiple-nanoseconds).\\

\noindent
Overall, the present work finds that s-CI and l-CI can reside on top
or within water droplets and their spectroscopy depends somewhat on
the energy function used. With a MS-ARMD parametrization the spectral
lines considered all remain unshifted or shift to the blue whereas the
ML-PES leads to both, blue- and red-shifted features. One important
difference between the two representations is that MS-ARMD is based on
fixed point charges whereas the ML-PES (using the PhysNet
architecture) includes geometry-dependent charges and it is
anticipated that fluctuating charge models are better suited for
quantitative studies. Importantly, using both PESs the magnitude of
the spectral shifts between the gas-phase and in/on-droplet are
consistent with typical Stark-induced red shifts from experiments and
simulations in the condensed phase.\\

\noindent
In conclusion, the spectroscopy of s-CI and l-CI in/on droplets and on
ASW differs in characteristic ways from the gas-phase spectra but for
the droplet sizes considered here, no differences for ``inside'' and
``on surface'' positions for droplets are found. On ASW the
diffusional barriers are sufficiently high at low temperatures (50 K)
to prevent surface diffusion whereas for the droplets facile migration
between ``inside'' and ``on surface'' positions is observed.\\

\section*{Acknowledgment}
This work has been financially supported by the Swiss National Science
Foundation (200021-117810, 200020-188724), and the University of
Basel.\\

\section*{Supporting Information}
The supporting material includes Tables with frequencies, additional
radial distribution functions and comparisons between spectra for
``inside'' and ``on surface'' positions of s-CI and l-CI.

\section*{Data Availability}
Relevant data for the present study are available at
\url{https://github.com/MMunibas/environment/}.

\clearpage
\bibliography{refs}
\clearpage

\renewcommand{\thetable}{S\arabic{table}}
\renewcommand{\thefigure}{S\arabic{figure}}
\renewcommand{\thesection}{S\arabic{section}}
\renewcommand{\d}{\text{d}}
\setcounter{figure}{0}  
\setcounter{section}{0}  
\setcounter{table}{0}

\newpage

\noindent
{\bf SUPPORTING INFORMATION: Structure and Spectroscopy of Criegee
  Intermediates in Gas- and Aqueous Environments}

\section*{Tables}
\begin{table} [H]
    \centering
    \begin{tabular}{c|c|c|c|c}
    \hline
    Normal Modes & CASPT2/aVTZ & CCSD(T)-F12a/aVTZ & PhysNet & Exp\cite{lee:2013}\\
    \hline
    $\nu_1$: asymmetric C–H stretch & 3328.37 & 3304.02 & 3333.66\\
    $\nu_2$: symmetric C–H stretch & 3164.23 & 3144.62 & 3169.95\\
    $\nu_3$: H-C-H scissor & 1487.95 & 1492.57 & 1486.44 & 1435\\
    $\nu_4$: C-O stretch/H-C-H scissor & 1273.88 & 1318.64 & 1285.72 & 1286\\
    $\nu_5$: C–O stretch & 1239.27 & 1244.46 & 1241.02 & 1241\\
    $\nu_6$: O–O stretch & 1001.58 & 952.36 & 966.13 & 908\\
    $\nu_7$: H-C-H-rock & 748.18 & 873.02 & 816.99 & 848\\
    $\nu_8$: H-C-H-wag & 630.72 & 650.04 & 612.74\\
    $\nu_9$: C-O-O scissor & 543.26 & 539.67 & 554.63\\
    \hline
    \end{tabular}
    \caption{Harmonic frequencies of H$_2$COO from the {\it ab initio}
      calculation, the resent PhysNet model, and
      experiment\cite{lee:2013}, in cm$^{-1}$.}
    \label{sitab:frequencies_h2coo}
\end{table}

\begin{table} [H]
    \centering
    \begin{tabular}{c|c|c|c|c}
    \hline
    Normal Modes & MP2/aVTZ & MP2/6-311++G(2d,2p) & PhysNet & Exp\cite{lee:2015}\\
    \hline
    $\nu_1$ & 3253.22 & 3286.83 & 3251.91\\
    $\nu_2$ & 3208.70 & 3240.90 & 3207.67\\
    $\nu_3$ & 3101.20 & 3146.18 & 3102.08\\
    $\nu_4$ & 3047.81 & 3088.37 & 3048.12\\
    $\nu_5$ & 1514.80 & 1531.36 & 1513.69 & 1477\\
    $\nu_6$ & 1474.35 & 1522.91 & 1474.24\\
    $\nu_7$ & 1456.37 & 1500.14 & 1457.16\\
    $\nu_8$ & 1397.69 & 1446.74 & 1397.50\\
    $\nu_9$ & 1296.20 & 1298.27 & 1295.50 & 1281\\
    $\nu_{10}$ & 1130.54 & 1163.85 & 1129.79 & 1091\\
    $\nu_{11}$ & 1031.22 & 1052.81 & 1030.95\\
    $\nu_{12}$ & 996.49 & 1013.63 & 995.86 & 956\\
    $\nu_{13}$ & 940.31 & 958.16 & 939.79 & 871\\
    $\nu_{14}$ & 745.36 & 721.75 & 744.90\\
    $\nu_{15}$ & 698.64 & 709.42 & 698.19\\
    $\nu_{16}$ & 480.98 & 470.49 & 481.11\\
    $\nu_{17}$ & 303.96 & 310.34 & 308.88\\
    $\nu_{18}$ & 224.63 & 216.80 & 223.27\\
    \hline
    \end{tabular}
    \caption{Harmonic frequencies of {\it syn}-CH$_3$CHOO from the
      {\it ab initio} calculation and the available experiment
      measurement,\cite{lee:2015} in cm$^{-1}$.}
    \label{sitab:frequencies_ch3choo}
\end{table}

\begin{figure} [H]
    \centering
    \includegraphics[width=1.0\linewidth]{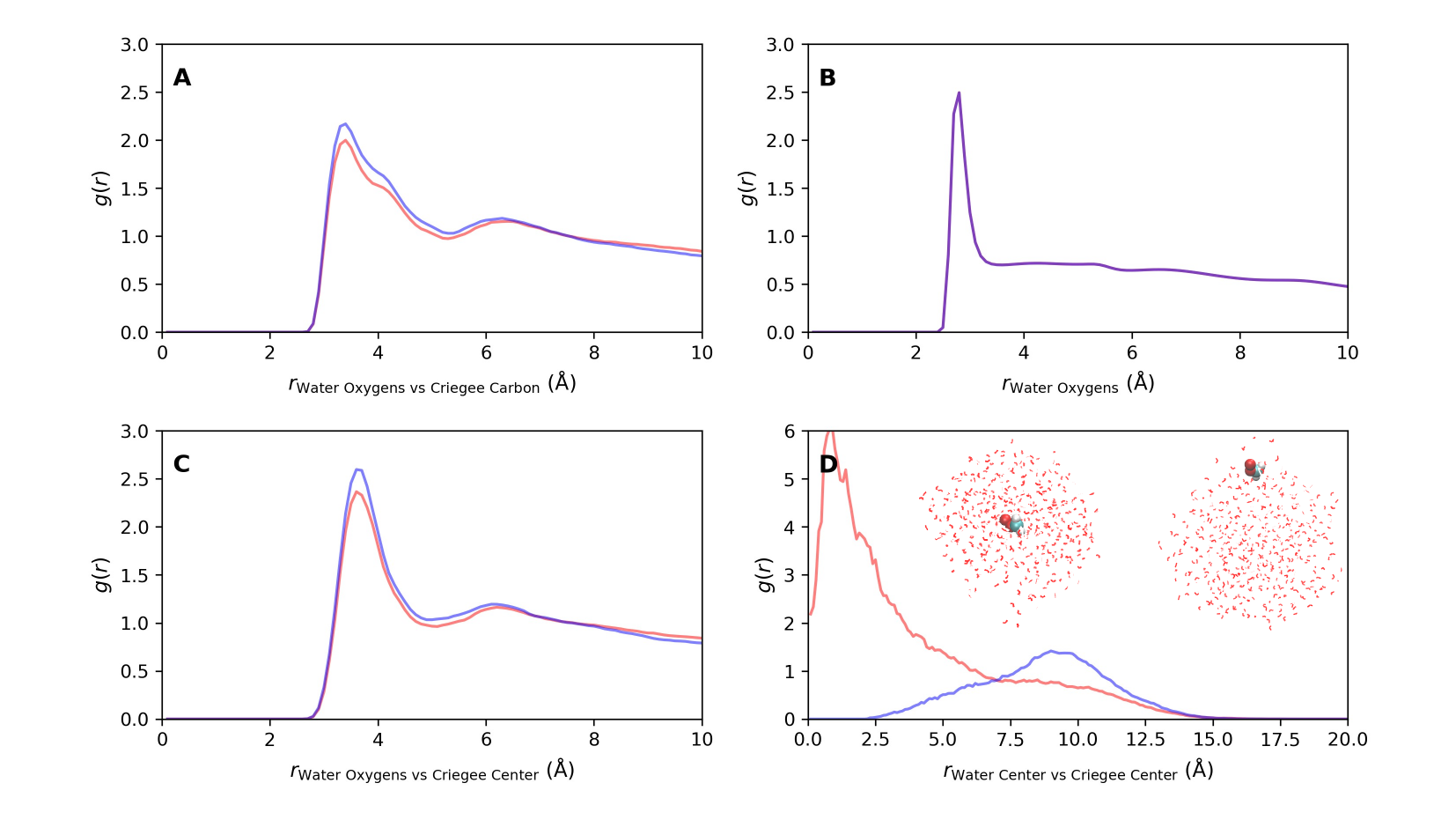}
    \caption{Radial pair distribution function for H$_2$COO (CI) on
      the water droplets. The radial distances in panels A to D are
      the O$_{\rm W}$--C$_{\rm CI}$, O$_{\rm W}$--O$_{\rm W}$, O$_{\rm
        W}$--CoM$_{\rm CI}$, and CoM$_{\rm droplet}$--CoM$_{\rm CI}$
      separations. The data from MS-ARMD and PhysNet simulations are
      represented by red and blue curves, respectively. The insets
      provide snapshots illustrating the CI located inside the water
      droplet and adsorbed on its surface.}
    \label{sifig:droplet}
\end{figure}

\begin{figure} [H]
    \centering
    \includegraphics[width=1.0\linewidth]{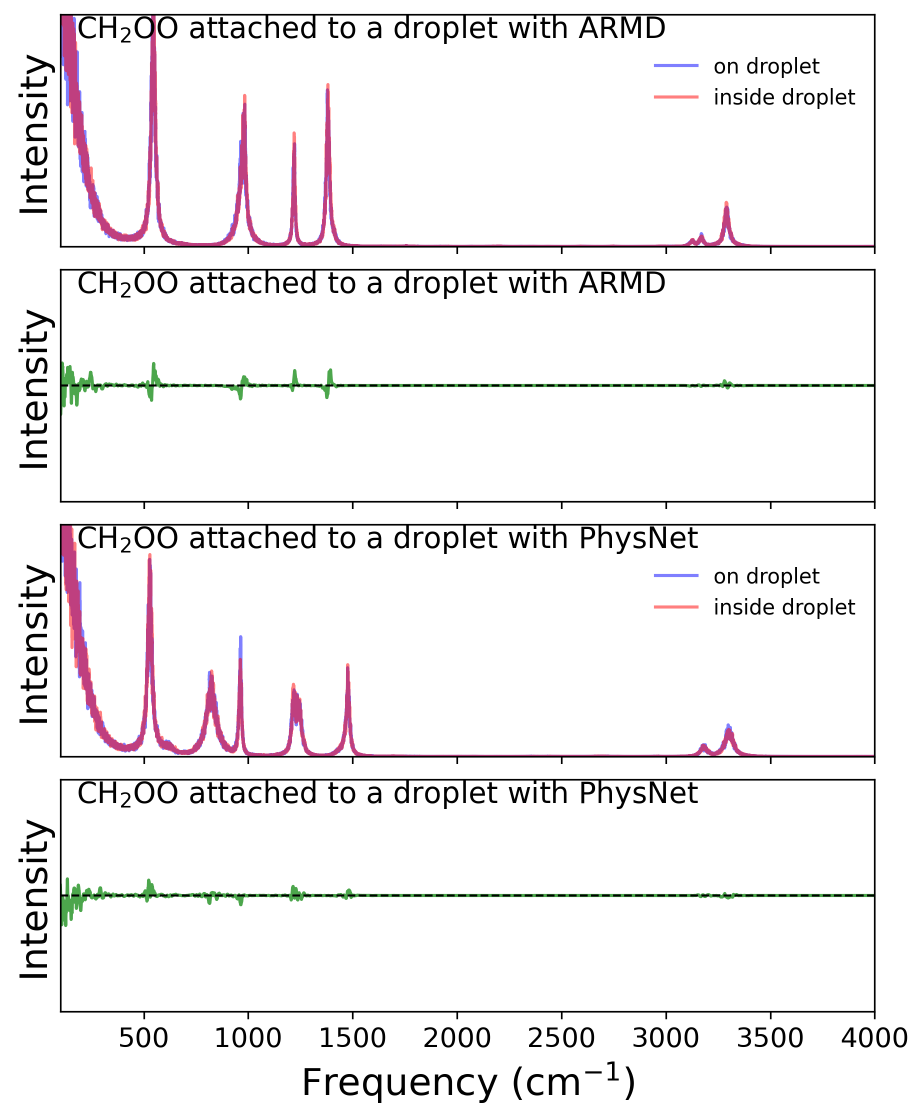}
    \caption{Comparison of IR spectrum of H$_2$COO between Criegee on
      the surface .vs. inside a droplet. The differences between the
      two cases are shown below.}
    \label{sifig:c1_on_in_full}
\end{figure}

\begin{figure} [H]
    \centering
    \includegraphics[width=1.0\linewidth]{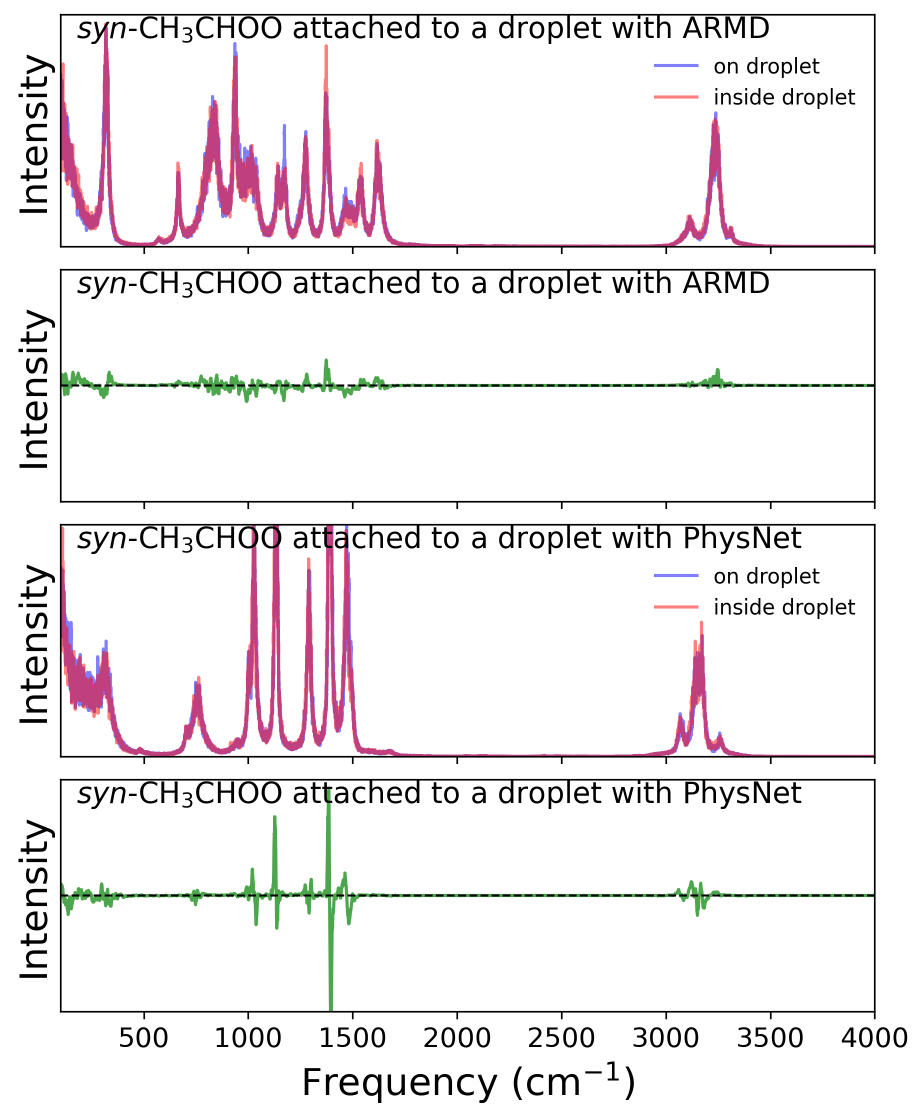}
    \caption{Comparison of IR spectrum of {\it syn}-CH$_3$CHOO between
      Criegee on the surface .vs. inside a droplet. The differences
      between the two cases are shown below.}
    \label{sifig:c2_on_in_full}
\end{figure}

\end{document}